\documentclass[aps,prl,twocolumn,showpacs, superscriptaddress]{revtex4-1}
\usepackage{graphicx}% Include figure files
\usepackage{dcolumn}% Align table columns on decimal point
\usepackage{bm}% bold math
\usepackage{epstopdf}
\usepackage{amsmath}
\usepackage{amstext} 
\usepackage{color}
\usepackage{tabularx}
\usepackage[normalem]{ulem}
\usepackage{hyperref}

\begin{document}

\definecolor{darkgreen}{rgb}{0.2,0.5,0.2}
\newcommand{\he}{$^3$He }
\newcommand{\xe}{$^{129}$Xe }
\newcommand{\hexe}{$^3$He-$^{129}$Xe}
\newcommand{\tighter}[1]{\textcolor{blue}{#1}}
\newcommand{\HL}[1]{\textbf{\textcolor{darkgreen}{(#1)}}}
\newcommand{\WT}[1]{\textbf{\textcolor{blue}{(#1 -- WT)}}}
\newcommand{\w}{$\omega$}
\newcommand{\m}{$\mu$}
\newcommand{\ke}{$\kappa_{\mathrm{el}}$}
\newcommand{\kg}{$\kappa_{\mathrm{geom}}$}
\newcommand{\mf}[1]{\mathrm{#1}}
\newcommand{\dg}{$^{\circ}$}
\newcommand{\fc}{$\widetilde{\omega}(t)\,$}

\newcommand{\mlk}{$M_k^{\mf{L}}$}
\newcommand{\mlm}{$M_m^{\mf{L}}$}

\newcommand{\mti}{$M_i^{\mf{T}}$}
\newcommand{\ml}{$\mf{M^{L}}$}
\newcommand{\mt}{$M^{\mf{T}}$}
\newcommand{\mlkm}{$M^{\mf{L}}_{k,m}$}
\newcommand{\Bint}{ {\bf B}$_\mf{int}$}

\title{Frequency shifts in noble-gas comagnetometers \\ \today}

\author{W.\,A.~Terrano}
\altaffiliation{Present Address: Department of Physics, Princeton University, Princeton NJ 08550 USA}
\email{wterrano@princeton.edu}

\affiliation{Physikdepartment, Technische Universit\"{a}t M\"{u}nchen, Boltzmannstr. 2 / EXC, 85748 Garching, Germany.}

\author{J.~Meinel}
\altaffiliation{Present Address: University of Stuttgart, 3rd Physics Institute, Pfaffenwaldring 57, Stuttgart 70569, Germany}
\email{jonas.meinel@pi3.uni-stuttgart.de}

\affiliation{Physikdepartment, Technische Universit\"{a}t M\"{u}nchen, Boltzmannstr. 2 / EXC, 85748 Garching, Germany.}
\author{N.~Sachdeva}
\affiliation{Department of Physics, University of Michigan, Ann Arbor, Michigan 48109, USA}
\author{T.E.~Chupp}
\affiliation{Department of Physics, University of Michigan, Ann Arbor, Michigan 48109, USA}
\author{S.~Degenkolb}
\affiliation{Institut Laue-Langevin, CS 20156 F-38042 Grenoble Cedex 9, France}
\author{P.~Fierlinger}
\affiliation{Physikdepartment, Technische Universit\"{a}t M\"{u}nchen, Boltzmannstr. 2 / EXC, 85748 Garching, Germany.}
\author{F.~Kuchler}
\affiliation{TRIUMF, Vancouver, BC V6R 2Z9, Canada}
\author{J.T.~Singh}
\affiliation{National Superconducting Cyclotron Laboratory and Department of Physics \& Astronomy, Michigan State University} 

\begin{abstract}

Polarized nuclei are a powerful tool in nuclear spin studies and in searches for beyond-the-standard
model physics. 
Noble-gas comagnetometer systems, which compare two nuclear species, have thus far been limited by anomalous frequency variations of unknown origin.
 We studied the self-interactions in a
 $^3$He-$^{129}$Xe system by independently addressing, controlling and measuring the influence of each
component of the nuclear spin polarization. Our results directly rule out prior explanations of the shifts,  and demonstrate experimentally that they can be explained by species dependent self-interactions. We also report the first gas phase frequency shift
induced by \xe on $^3$He. 
\end{abstract}
\pacs{32.30.Dx, 06.30.Gv}

\maketitle

Noble gas NMR techniques \cite{Chupp1988}  find applications in medical
imaging \cite{Moeller2002, Meersman2015}, atomic gyroscopes \cite{Kornack2005} and tests of beyond
the standard model physics \cite{Vasilakis2009, Brown2010, Smiciklas2011, Lee2018}.  
The most precise applications are often limited by hitherto
unaccounted for anomalous frequency variations.
Understanding
the physical origin of these variations directly impacts
the future of $^3$He-$^{129}$Xe probes for Lorentz-violation \cite{Bear2000, Gemmel2010, Allmendinger2014}, the \xe electric dipole moment \cite{Rosenberry2001}, fifth
forces \cite{Tullney2013}, and direct-detection of axionic and ``Fuzzy''
dark matter \cite{Graham2018}. More generally, some types of precision
atomic gyroscopes \cite{Limes2018b}, magnetometers \cite{Koch2015} and, possibly,
quantum memory technologies \cite{Katz2018} will need to account
for these effects. 

Fully exploiting
the sensitivity of these techniques requires understanding
the self-interactions of the gases, as was made
clear by a recent test of Lorentz violation using a cohabitating
$^3$He-$^{129}$Xe magnetometer \cite{Allmendinger2014}. That work set
a limit on preferred reference frames
in the nuclear sector that remains the tightest by a factor of four. The limit of 3.6 nHz on sidereal frequency variations was extracted on top of $\mu$Hz-level anomalous frequency variations.
The explanation for these variations in terms of self-interactions due to the transverse gas
magnetization was controversial \cite{Romalis2014, Allmendinger2014b}, and as demonstrated here, incorrect. In
this paper we present a new technique that allows dynamic control of each component of the nuclear magnetization
and use it measure the self-interactions of the $^3$He-$^{129}$Xe system. Our results rule
out transverse-magnetization as the dominant source of the frequency variations
and show that self-interactions coupling to the longitudinal
magnetization can explain the observed variations. 

A comagnetometer experiment corrects for the effects of magnetic field variations by comparing the frequencies or phases of two species, for instance by defining the corrected frequency 
\begin{equation}\label{eq: co}
\widetilde{\omega}_{k}(t) = \omega_{k}(t) - \omega_{m}(t) \gamma_{k}/\gamma_{m}, 
\end{equation}

where $k$ and $m$ label the two distinct spin species and $\gamma_{k},\gamma_{m}$ are their gyromagnetic ratios. 
Searches for new physics look for variations in \fc that correlate with an experimental parameter. 

\begin{figure}[htp]
\scalebox{0.23}{\includegraphics{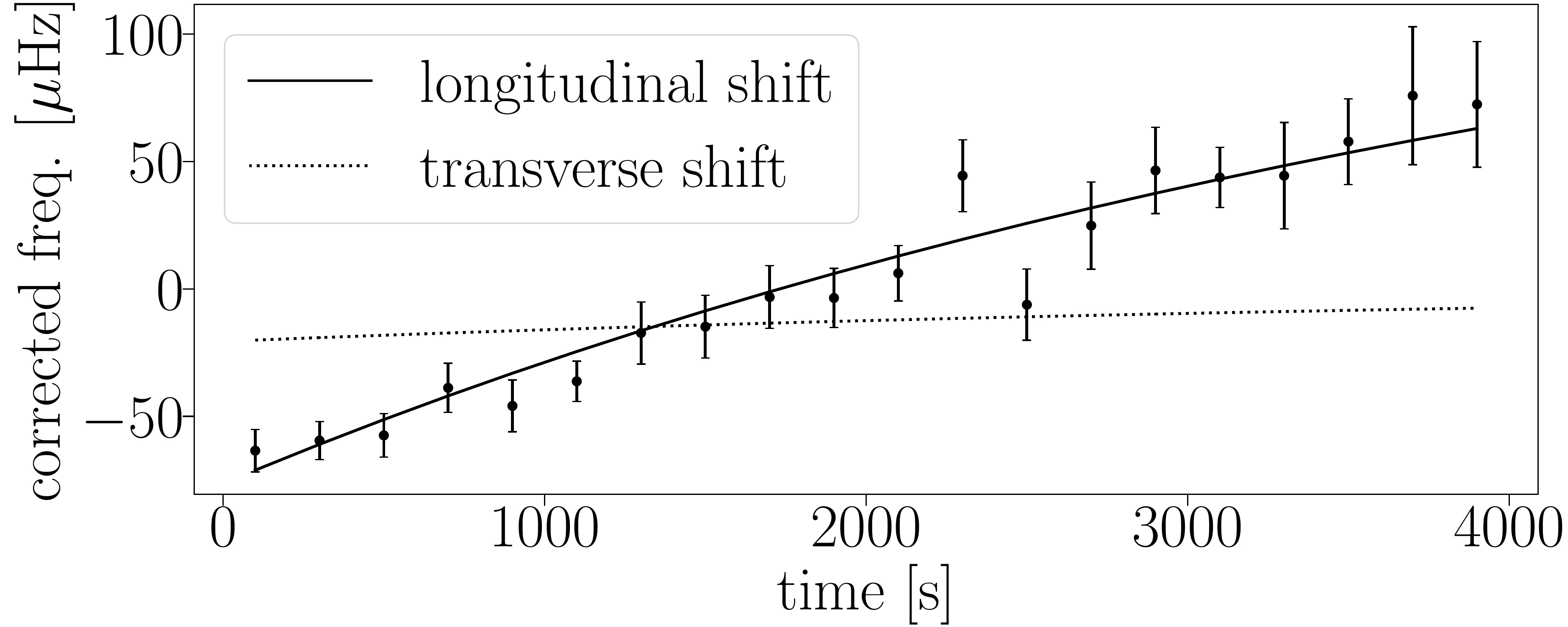}}
\caption{
The corrected frequency $\widetilde{\omega}_{\mf{He}}(t)/2\pi$ (Eq. \ref{eq: co}, offset subtracted) from our data, showing variations similar to those previously reported.  
The dotted curve shows the largest shift due to transverse magnetization that is consistent with the results of this paper, showing that the previously proposed explanation for the variations is excluded.
Shifts proportional to longitudinal magnetization, also measured here, match the observed variations well (solid curve). See \emph{Summary and conclusions} for details of the models.
}
\label{fig: drift fit} 

\end{figure}

Several recent experiments  reported anomalous variations in \fc on time scales of several hundred seconds\cite{Rosenberry2001, Gemmel2010, Gemmel2010b, Allmendinger2014, Tullney2013}. %
We observed similar variations in our apparatus, Figure \ref{fig: drift fit} shows a representative example. 

To investigate the source of comagnetometer variations, we directly measured frequency shifts proportional to the transverse-rotating ($M^\mf{T}\propto\sin\theta_\mf{s}$) and longitudinal-static ($M^\mf{L}\propto\cos\theta_\mf{s}$) magnetizations of each species.  Here $\theta_\mf{s}$ is the tip angle of the spins relative to the background magnetic field.   We characterize the shifts in $\omega_k$ in terms of coupling parameters $\rho$ and $\lambda$:

\begin{subequations}
\begin{tabular}{lc}
\parbox{3.8cm}
{\begin{equation} \omega_{k}^{\mf{T}}=\!\!\!\!\sum\limits_{i=\mf{He, Xe}}\!\!\rho_k^i M^\mf{T}_i  \label{eq: trans}
\end{equation}}&
\parbox{3.9cm}{\begin{equation}
\omega_{k}^{\mf{L}}= \!\!\!\!\sum\limits_{i=\mf{He, Xe}}\!\!\lambda_k^i M^\mf{L}_i \label{eq: long}
\end{equation} }
\end{tabular}
\end{subequations}
which can produce time-dependent drifts in the corrected frequency

\begin{align}
\widetilde{\omega}^\mf{T}_{k}(t) = \widetilde{\rho}^{\,k} M_k^\mf{T}(0) e^{-t/{T_{2}^*}^{\!(\!k\!)}} - r_{km} \widetilde{\rho}^{\,m} M_m^\mf{T}(0) e^{-t/{T_{2}^*}^{\!(\!m\!)}}  \label{eq: trans drift}\\
\widetilde{\omega}^\mf{L}_{k}(t) = \widetilde{\lambda}^{k} M_k^\mf{L}(0) e^{-t/T_1^k} - r_{km} \widetilde{\lambda}^m M_m^\mf{L}(0) e^{-t/T_1^m} \label{eq: long drift}
\end{align}
as the gas magnetizations $M^\mf{T}$ and $M^\mf{L}$ decay with phenomenological time-constants $T_{2}^*$ and $T_1$. Here $r_{km} = \gamma_k/\gamma_m$, $\widetilde{\lambda}^k = \lambda^k_k -  r_{km} \lambda^k_m$ and  $\widetilde{\rho}^{\,k} = \rho^k_k - r_{km} \rho^k_m$. 

Our main findings are: 
(i)
 transverse frequency shifts (Eq. \ref{eq: trans}) cannot explain the variations we measured in $\widetilde{\omega}(t)$, contradicting several prior papers \cite{Gemmel2010, Tullney2013, Allmendinger2014}
(ii) longitudinal frequency shifts (Eq. \ref{eq: long}) are the largest effect and $ \lambda^k_k /\gamma^k \neq \lambda^k_m /\gamma^m$ so, crucially, the longitudinal shifts do not cancel in the comagnetometer and can produce slow frequency variations and (iii) the longitudinal comagnetometer shift is due to resonant effects and direct contact interactions between the noble-gas nuclei rather than magnetic-gradient sampling effects.
 
\noindent{\emph{Parametrizations, and theory of internal fields --- }}

A key point of controversy \cite{Allmendinger2014, Romalis2014, Allmendinger2014b, Rosenberry2000, Oteiza1992}
has been the magnitude of the internal magnetic fields ($\mathbf{B}_{\mf{int}}$) in a Rb-free \hexe\ cell.  Inside a uniformly magnetized sphere the field experienced by species $k$ is entirely due to contact interactions with species $m$.  This gives $\mathbf{B}_{\mf{int}, m} = \frac{2\mu_0}{3} \kappa_{km} \mathbf{M}_{k}$ where $\kappa$ parameterizes the overlap between the spin-species \cite{Schaefer1989kappa}.  This is a scalar interaction and symmetric for $k\leftrightarrow m$.
 Since the \he and \xe nuclei do not directly overlap, $\kappa_{km}$ is zero to first-order for a \hexe\ gas mixture.  Contact interactions require higher-order couplings through the electronic spins or a mediator species \cite{Vlassenbroek1996}.  Recently, a non-zero $\kappa_\mf{HeXe}$ was measured in a \hexe\ comagnetometer with a cohabitating Rb read-out \cite{Limes2018b}. 
 
Deviations from a spherical geometry produce long-range dipolar fields that do not average to zero. We parametrize these fields in terms of $B_\mf{dip}^{i} = \mu_0 \Gamma^{i} M^{i}$ where $ \Gamma^{i}$ are dimensionless geometric factors.
 
Internal fields from the precessing nuclei can apply Ramsey-Bloch-Siegert shifts to the other nuclei \cite{Gemmel2010}.  This is the basis for previous explanations that claimed that the transverse magnetization was the origin of the frequency variations.

The relaxation-free Bloch equations $\mf{d}{\bf M}_k/\mf{d}t = \gamma_k {\bf M}_k \times \bf{B}_\mf{int}$ taken together with ${\bf B}_\mf{int} = B^i_\mf{dip} = \mu_0 \Gamma^i M^i$, show a resonant shift due to the longitudinal magnetization ($M^{\mf{L}}$) in a non-spherical cell.
Averaging over a Larmor cycle, the transverse field becomes $B_k^{\mf{T}} = \mu_0 \Gamma^{\mf{T}} M_k^\mf{T}$, and the transverse magnetization precesses at 
\begin{equation}
\label{eq: bloch}
\omega_k =\mu_0 \gamma_k \big((\Gamma^\mf{L} - \Gamma^\mf{T}) M_k^\mf{L} + \Gamma^\mf{L} M_m^\mf{L}\big)
\end{equation}
relative to the frame rotating at $\gamma_k \mathbf{B}_0$, where $\mathbf{B}_0$ is the external holding field.
If the cell is not symmetric about  $\mathbf{B}_0$, the variations in $\Gamma$ add harmonics to Eq. \ref{eq: bloch}.
The $\Gamma^\mf{L}$ terms are the net field produced by the longitudinal gas polarizations, and cancel in $\widetilde{\omega}$.  The $\Gamma^\mf{T}$ term is an additional shift that does not cancel in $\widetilde{\omega}$ as it arises from the resonant torque produced by $M_k^\mf{T}$ on $M_k^\mf{L}$. 

Physically, the transverse internal field $B_k^{\mf{T}}$ is resonant with $M^{\mf{L}}_k$ and rotates $M^{\mf{L}}_k$ into the transverse plane at 90$^\circ$ to the existing transverse magnetization $M^{\mf{T}}_k$.  This causes the transverse magnetization of $k$ to change orientation. In contrast, $M^{\mf{L}}_m$ is not resonant with $B_k^{\mf{T}}$ and experiences no such effect.  

Contact interactions (which produce only heteronuclear shifts) and the resonant effects of Eq. \ref{eq: bloch} (which produce only homonuclear shifts) both affect the corrected frequency.  The combined effects are

\begin{equation}
\begin{split}
\label{eq: error}
\frac{\omega_k}{\mu_0 \gamma_k} &= (\Gamma^\mf{L} - \Gamma^\mf{T}) M^\mf{L}_k + (\Gamma^\mf{L} + 2\kappa_{km}/3) M^\mf{L}_m \\
\frac{\widetilde{\omega}_{k}}{\mu_0 \gamma_k} &= \Gamma^{\mf{T}} (M_{\mf{}m}^{\mf{L}} - M_{\mf{}k}^{\mf{L}}) + 2(\kappa_{km}M_{\mf{}m}^{\mf{L}} - \kappa_{mk}M_{\mf{}k}^{\mf{L}})/3.
\end{split}
\end{equation}

Independent control of the two species allows us to separately measure each term in Eq. \ref{eq: error}. 

\emph{Apparatus and data reduction --- }

\begin{figure}[h]
%\begin{center}
\scalebox{.22}{\includegraphics{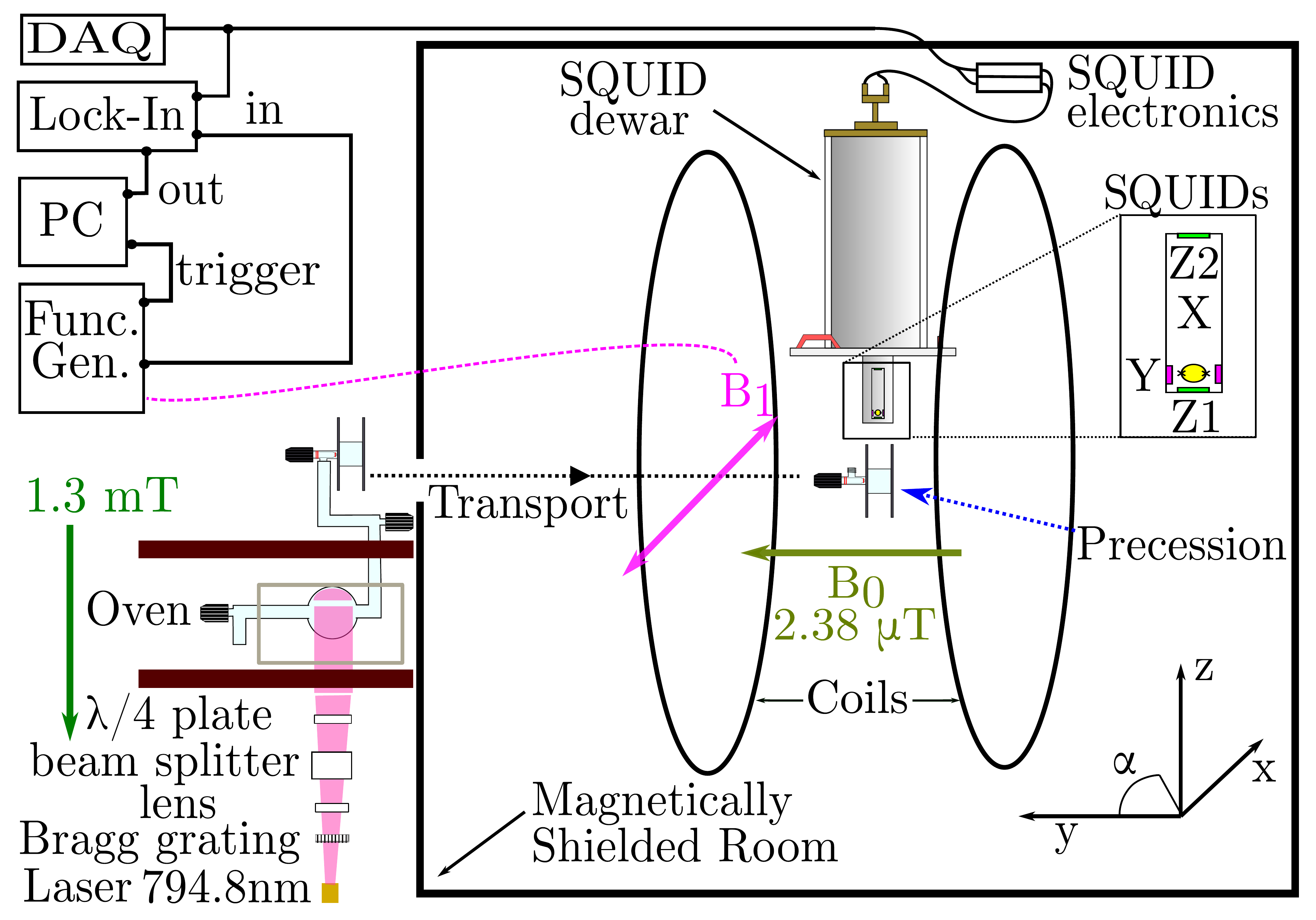}}
\caption{Diagram of our apparatus.  The gas was polarized  outside the magnetically shielded room; the spin-precession measurements took place inside the room, directly beneath the SQUID magnetometer system. The lock-in/PC/function-generator system allowed us to change the tip angles of the magnetizations during a run.}
\label{fig: expt} 
%\end{center}
\end{figure}

Figure \ref{fig: expt} shows a diagram of the experiment at the FRM-II in Munich.

We used two measurement cells: a sealed cell containing Rb and about 0.5 bar of $^3$He for single-species studies, and a valved cell filled with pre-polarized $^3$He-$^{129}$Xe-N$_2$ gas mixture at pressures ranging from 0.3 to 1.6 bars for dual-species studies. %
The cells were made from 2\,mm thick GE-180 glass.  The sealed cell was a blown sphere with a 33\,mm outer diameter (OD) bulb and a 27\,mm long  by 6.2\,mm OD pull-off stem. 
The valved cell was a 24.8\,mm long, 21.2\,mm OD cylinder bonded to doped-Si wafer end-caps.   The valve sealed a small hole in the center of one wafer.

Large \he and \xe polarizations were generated by spin-exchange optical pumping using the 794.8\,nm D$_1$ line of a Rb vapor \cite{Walker1997, Gentile2017}. 
Polarizing the \he took several hours at 150\dg C, while polarizing the \xe took 10 minutes at 110\dg C due to its larger spin-exchange rate. We then cooled the cell and adiabatically transported it into the magnetically shielded room where the measurement took place at 28\dg C \cite{Kuchler2016}.

A 1.6\,m diameter $y$-axis Helmholtz coil provided a 2.38\,$\mu$T holding field (${\bf B}_{0}$). Resonant fields (${\bf B}_{1}$) applied at 77.2\,Hz and 28.0\,Hz with a 1.5\,m diameter $x$-axis Helmholtz coil changed the precession tip angle of the \he and \xe spins, respectively.
A set of six SQUID magnetometers directly above the measurement cell
 monitored the precession of the \mt \,components of the gases.\
The SQUID system \cite{Drung2002} (lent by PTB-Berlin) contained two SQUIDs oriented along each axis.   

Subtracting the signals from the two z-axis SQUID magnetometers (separated by 12 cm) formed a gradiometer signal Z$_{\mf{grad}} $ to suppress background magnetic field fluctuations. The center of the measurement cell was situated variously between 2.8\,cm and 5.8\,cm below the lower SQUID.
 
Changing the tip angle of the precessing spins required ${\bf B}_1$ pulses with a particular phase relative to the spins.  In order to control for phase drifts between the clock and the spins we triggered the  ${\bf B}_1$  pulses from the Z$_1$ SQUID output.

We recorded the SQUID output signals at a sampling rate of 5\,kHz using a 24-bit digitizer (D-TACQ), which was stabilized by an atomic clock (SRS FS725).  After downsampling the data to 500\,Hz, we divided it into 5~second sections and fitted each section $n$ of Z$_{\mf{grad}} $  to
\begin{multline}
a_{\mf{He}} \sin(\omega_{\mf{He}} t) +  b_{\mf{He}} \cos(\omega_{\mf{He}} t) \\ + a_{\mf{Xe}} \sin(\omega_{\mf{Xe}} t) +  b_{\mf{Xe}} \cos(\omega_{\mf{Xe}} t) + c_1 t + c_0
\label{eq: fit}
\end{multline}
where the $a, b, \omega$ and $c$ were free parameters.
$\arctan(a_{\mf{He,Xe}}/b_{\mf{He,Xe}}) = \phi^n_{\mf{He,Xe}}$ gave the instantaneous \he and \xe phases $\phi^n$ at the start of section $m$ (time $t^n$).  The total phase accumulated at $t^n$ was $\Phi^n_{\mf{He,Xe}} = \phi^n_{\mf{He,Xe}} + 2\pi N^n_{\mf{He,Xe}}$ where $N_{\mf{He,Xe}}$ counts the number of completed cycles.

To cancel magnetic field fluctuations we defined adjusted phases. For two species we used $\widetilde{\Phi}_k(t) = \Phi_k  - r_{km} \Phi_m$.  For single-species measurements we defined 
 $\widehat{\Phi}^n_k = \Phi_k(t^n) - \gamma_k \mathcal{G} \int_0^{t^n}{(B_y(t) - B_y(0)) dt}$, with $B_y$ measured by the y-axis SQUID magnetometers, which coincided with $\mathbf{B}_0$.  The scaling factor $\mathcal{G}$ was SQUID- and geometry-dependent. 
Fits to $\widetilde{\Phi}_k(t^n)$ or $\widehat{\Phi}_k(t^n)$  
gave the corrected frequencies and frequency variations.

\noindent\emph{Frequency shifts due to transverse magnetization --- } 

\begin{figure}[htp]
\scalebox{0.285}{\includegraphics{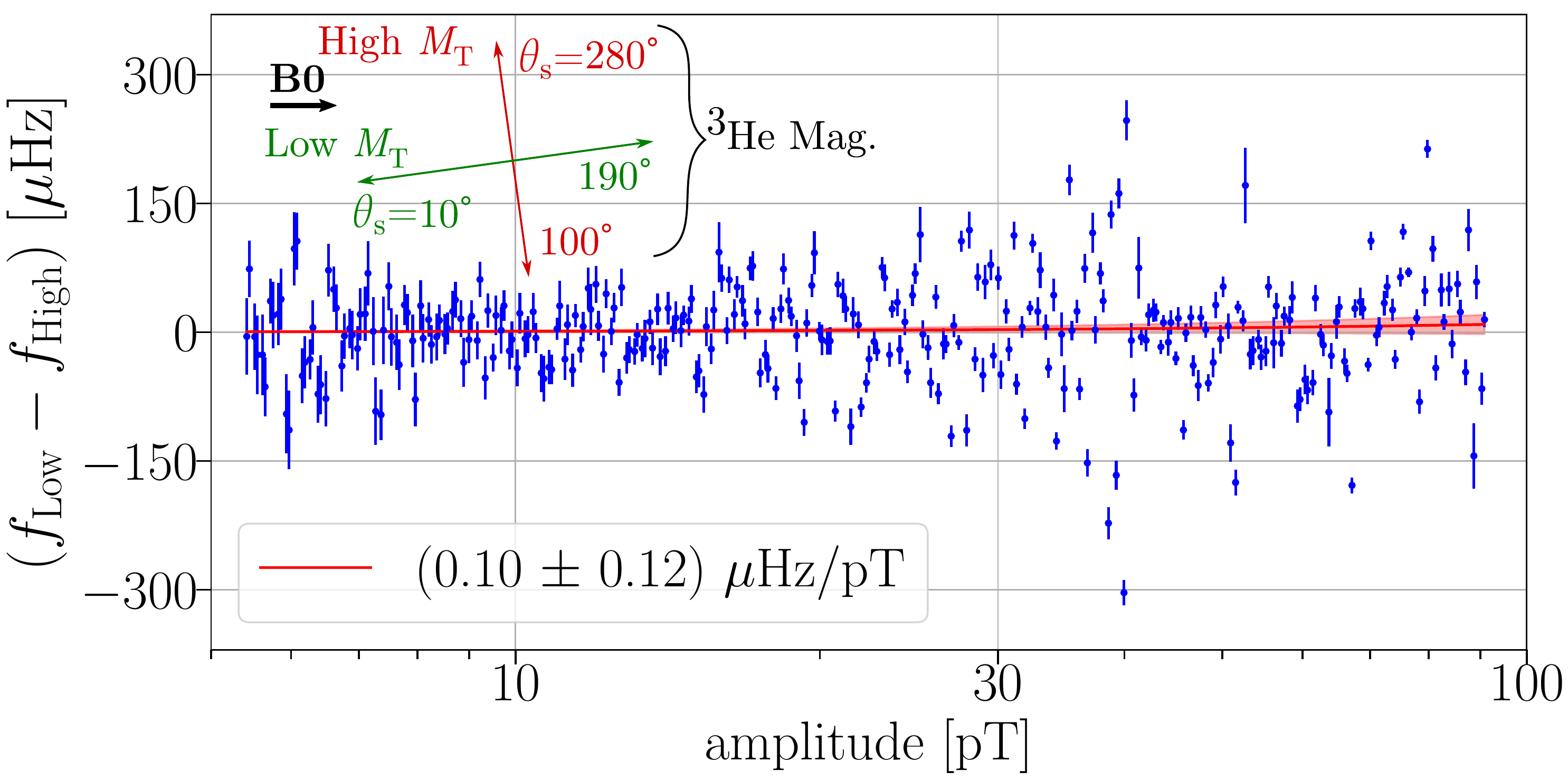}}
\caption{Difference in \he frequency at low ($f_\mf{Low}$) and high ($f_\mf{High}$) tip angles $\theta_\mf{s}$ as a function of field from the cell at the SQUID. Each point combines measurements with opposite $B_0$ projection. The increased scatter near 40 pT amplitude is due to large magnetic field drifts during those measurements. The slope  is consistent with zero.}
\label{fig: transverse} 
\end{figure}

The first explanations for the comagnetometer variations ascribed them to Ramsey-Bloch-Siegert shifts from the transverse magnetization of each species on itself \cite{Gemmel2010, Tullney2013, Allmendinger2014}.  Such effects are difficult to model, motivating direct experimental study.  An interaction of this type would result in a net shift between the frequency measured at low tip angle and at high tip angle.  Using the sealed cell, which could achieve very high \he magnetizations, we applied phase-matched NMR pulses to  move the magnetization between four tip angles: low (10\dg\ and 190\dg) and high (100\dg\ and 280\dg), as shown in Figure \ref{fig: transverse} inset.   Averaging pairs with opposite $\mathrm{B}_0$ projection cancels effects due to longitudinal magnetization.  

For every set of four tip angles we calculated the frequency difference between the high and low tip angle states and determined the transverse magnetization from the amplitude of the precession signal.  To cancel shifts due to magnetic drift  we reversed the tip angle sequence every 120 seconds.
 As shown in Figure \ref{fig: transverse}, we saw no evidence that the \he precession frequency depends on the magnitude of the transverse magnetization.  We constrain $\rho^\mf{He}_\mf{He}/2\pi < 6.1$\,mHz/(A/m) at the 68\% confidence level.

\noindent \emph{Frequency shifts due to longitudinal magnetization --- } 

While measuring transverse shifts we observed and canceled large longitudinal frequency shifts. 
To further investigate the longitudinal shifts, we applied a {\bf B}$_1$ produced both transverse and longitudinal magnetizations. A train of $\pi$-pulses then flipped $M\mf{^L}$ and any frequency shifts associated with it. 

\begin{figure}[h]
\scalebox{.23}{\includegraphics{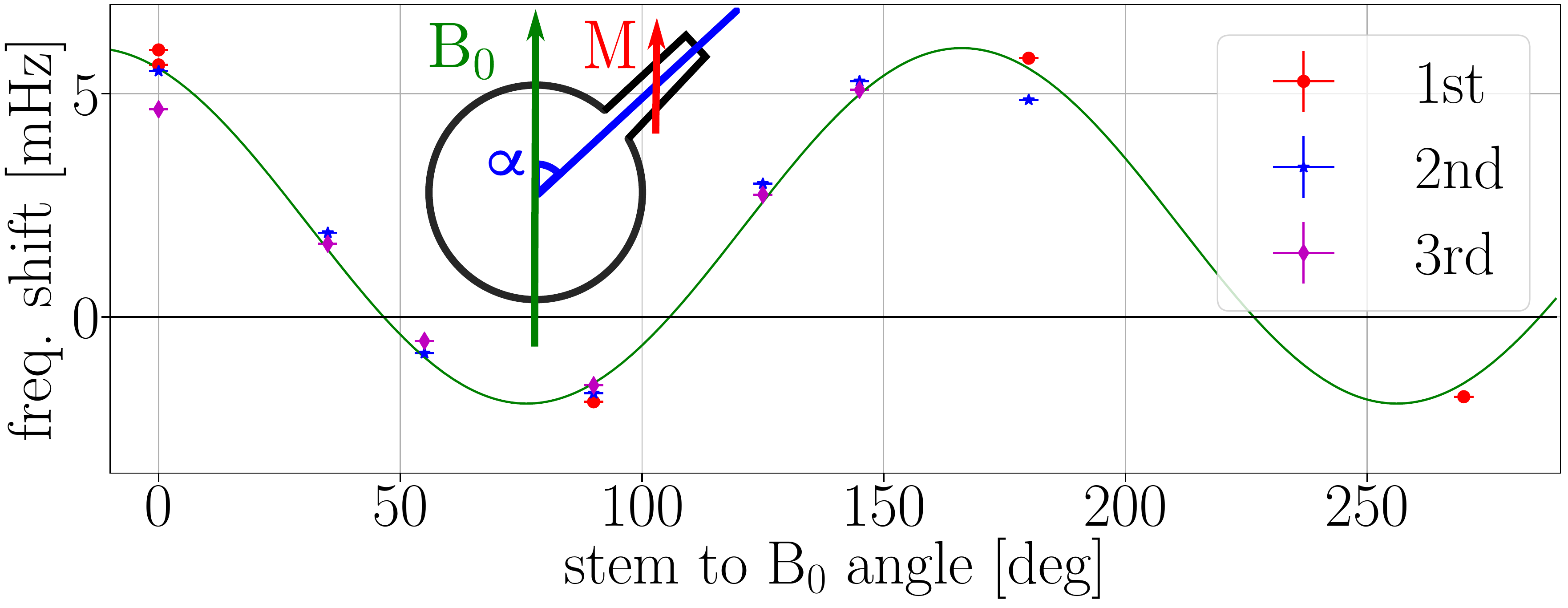}}
\caption{Change in \he frequency on inverting the \he magnetization, as a function of cell orientation.  The curve is $(3\cos^2{\alpha} - 1)\cdot2.7$ mHz, with a 0.7 mHz offset, corresponding to the angular dependence of the average field in the cell produced by a \he dipole (${\bf M}$) at the stem. The offset is likely due to $\alpha$-symmetric asphericities, such as oblateness of the sphere.}
\label{fig: stem angle}
\end{figure}

Figure \ref{fig: stem angle} shows the dependence of the longitudinal shift on cell orientation for the sealed cell.  The shift is proportional to $(3 A \cos^2 \alpha -A)$, where $\alpha$ is the stem-to-B$_0$ angle and $A$ is the shift amplitude, as is expected for a shift generated by the \he dipole in the stem. 
Based on analytical calculations using the measured cell geometry, we estimated that the gas in the stem would produce a net magnetic field of ($45\pm15$) pT across the cell, dominated by the field within the stem. The corresponding shift from a static dipole would be $A_\mf{dipole} = (1.5\pm0.5$) mHz.  The \he dipole, however, also has a rotating component, so the second term of Eq. \ref{eq: bloch} amplifies the frequency shift by a factor of 3/2 and we predict $A_\mf{model} = (2.25\pm0.75$) mHz. The measured $A_\mf{expt.} = (2.7\pm0.1$) mHz agrees with our geometric estimate of the resonant enhancement.

\begin{figure}[h]
\scalebox{.27}{\includegraphics{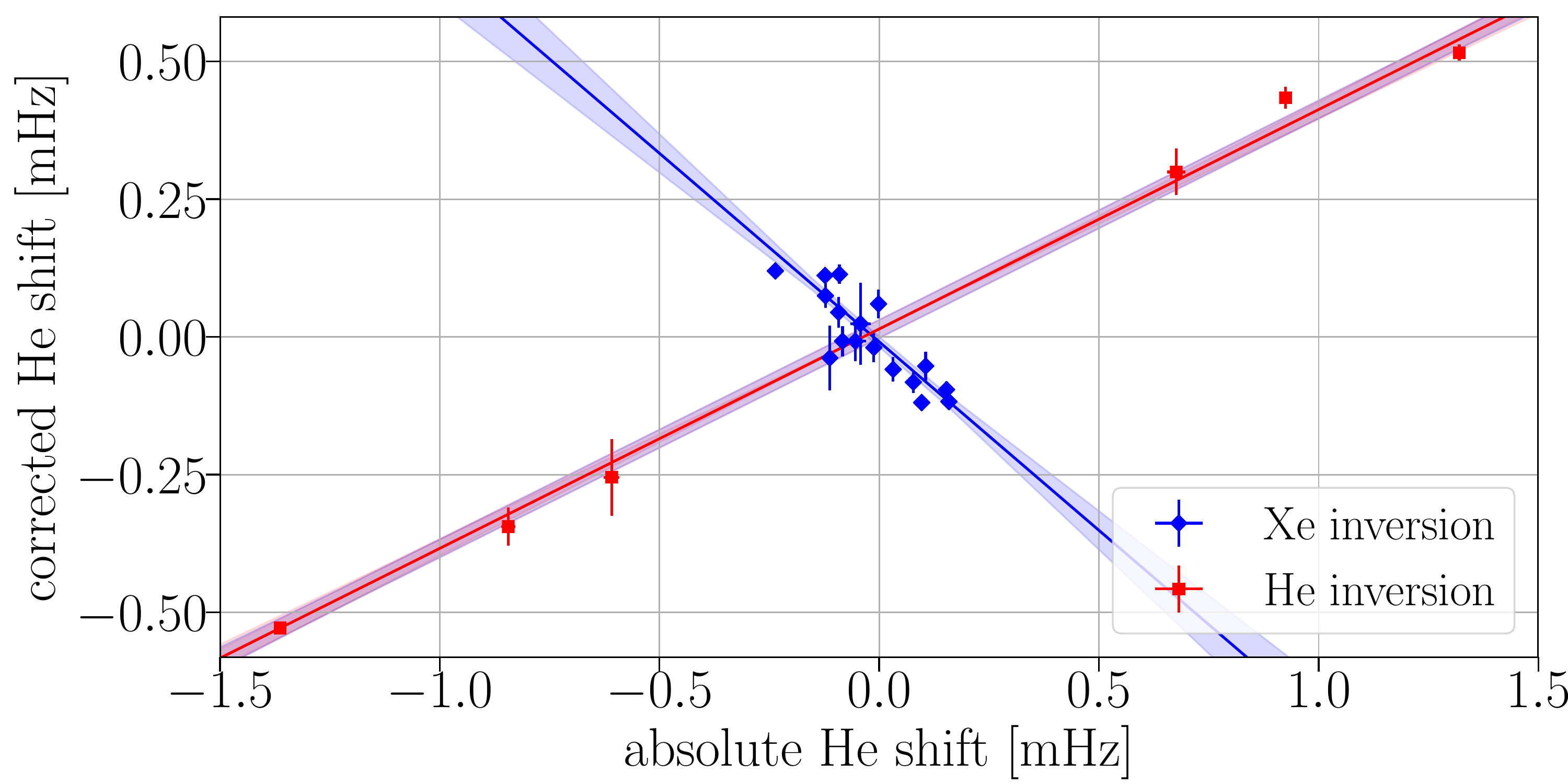}}
\caption{Change in the absolute ($\omega_{\mf{He}}$) and corrected ($\widetilde{\omega}_{\mf{He}}$) \he frequencies when the longitudinal magnetizations of \xe  and \he are inverted (blue diamonds and red squares).  Measurements taken in the valved cell, some errors are hidden by the symbols.  The slope of the lines measures the shifts in the ratios of interest, with 1-$\sigma$ error (shaded) from the covariance of the fit to a line.  If the comagnetometer correction canceled frequency shifts from longitudinal magnetization (Eq. \ref{eq: long}) the lines would be horizontal.}
\label{fig: correlation}
\end{figure}

Longitudinal shifts do not cancel in the corrected frequency, as shown in Figure \ref{fig: correlation}, so the decay of $M_m^\mf{L}$ causes time variations in $\widetilde{\omega}(t)$.   We experimentally measured $\widetilde{\lambda}^\mf{He} = (750\pm60$) mHz/(A/m) for our system, and isolated the physical mechanisms responsible for the finite $\widetilde{\lambda}$.

To investigate whether magnetic field gradients explained the non-zero $\widetilde{\lambda}$\cite{Sheng2014}, we used a small coil that mimicked the gradients of the cell.
For a given change in the Helium frequency, the corrected frequency shift produced by the coil is 100 times smaller than the shift produced by the nuclear-spin polarization.

To separately measure all of the geometric and contact interactions in Eq. \ref{eq: error}, we used each species as both source and probe of the longitudinal shifts, and changed the geometric effect by changing the cell orientation.  We analyzed in terms of $\Delta^{\!(m)} \mathcal{R}_k = \Delta^{\!(m)}(\widetilde{\omega}_{k}/\omega_k)$, the change in the frequency ratio of species $k$ when the longitudinal polarization of species $m$ is inverted.  This ratio is insensitive to changes in polarization and tip angle.
Doing this for all four combinations of $k$ and $m$ and at cylinder-axis-to-B$_0$ angles $\alpha = 0^\circ$ and $90^\circ$ gave us eight measurements with different sensitivities to $\kappa$ and $\Gamma$.  

Table \ref{tab: results} lists the measured shifts in $\Delta^{\!(m)} \mathcal{R}_k$, and the $\kappa$ values extracted from them.
 The data set consisted of six separate cell fillings at different pressures, with 16780 s of data and 130 shift measurements at $\alpha=0^\circ$ and 8660 s of data and 58 shift measurements at $\alpha=90^\circ$. To extract $\kappa$ from our data we used Eq. \ref{eq: error} along with the geometric relations of a dipole: $\Gamma^\mf{T}=-\Gamma^\mf{L}/2$, and $\Gamma^\mf{L}(90^\circ) = -\Gamma^\mf{L}(0^\circ)/2$, as a function of cell orientation $\alpha$. The uncorrected homonuclear shifts, combined with the measured amplitudes, magnetometer-cell distance and flip-angles gave $\Gamma^\mf{L}(0^\circ) = 0.046\pm0.004$.

We measured $\kappa_\mf{HeXe} = -0.0094\pm0.0004$, while Limes et al. recently measured $\kappa_\mf{HeXe} = -0.011\pm0.001$ \cite{Limes2018b}.  A first-principles electronic-structure calculation, performed following these initial reports of gas-phase interactions between nuclear spins, suggests that the  discrepancy is explained by the temperature dependence of the interaction \cite{Vaara2018}. Our data also gives the first measurement of the shift induced by \xe on $^3$He: $\kappa_\mf{XeHe} = -0.0072 \pm 0.0008$. The comparable sizes of $\kappa_\mf{HeXe}$ and $\kappa_\mf{XeHe}$ supports the scalar interaction picture for the frequency shifts.

Our measurements of the internal fields also constrain heteronuclear transverse shifts.  With typical values in our comagnetometer system $M^\mf{T}=8\cdot10^{-4}$A/m,  $\kappa \sim \Gamma \approx -0.01$ and $\Delta = (\omega_\mf{He} - \omega_\mf{Xe})/2\pi \sim 50$Hz, Ramsey-Bloch-Siegert shifts across species would be ($\gamma {\bf B}_\mf{int})^2/2 \Delta \sim 4\times 10^{-10}$\,Hz, far below the $\mu$Hz variations reported in \fc (Figure \ref{fig: drift fit} and Refs. \cite{Allmendinger2014, Allmendinger2014b}).

\emph{Summary and conclusions}

Figure 1 compares our measured $\widetilde{\omega}(t)$ with the maximum possible transverse shift consistent with our measurement $\rho^\mf{He}_\mf{He} < 6.1$ mHz/(A/m), taking $T_2^*$ and $M^\mf{T}$ from the  precession signal, and showing that transverse shifts are inconsistent with the observed drifts. 
The longitudinal shifts predicted by our measurements of $\widetilde{\lambda}^\mf{He} = 750$mHz/(A/m) are also shown, assuming typical values for our system of $T_1^\mf{Xe} =  3500$s, $T_1^\mf{He}$ =  5250s and $\theta_s = 89^\circ$; the model matches the data well for a wide range of T$_1$. 

Ruling out the previously published explanations for the drifts \cite{Gemmel2010, Tullney2013, Allmendinger2014, Allmendinger2014b} required a much better measurement of transverse shifts than have been performed for longitudinal shifts:  with $\theta_s \approx 90^\circ$, transverse shifts are significantly enhanced relative to longitudinal shifts. 
Still, we suggest that the \fc variation does in fact come from longitudinal shifts which do not cancel in the corrected frequency and which decay over the run.  
We directly measured the magnitude of such a shift, showed it is large enough to explain the drifts and showed it largely involves two mechanisms:  a resonant effect that rotates the longitudinal magnetization into the transverse plane and a direct \he-\xe scalar interaction. 

These undesirable variations in \hexe\ comagnetometers could be reduced by minimizing residual $M_\mf{L}$ 
and choosing cell geometries where the geometric and scalar internal shifts cancel \cite{Limes2018b}, giving \he\!-\xe comagnetometers a chance to live up to their potential.

\begin{acknowledgments}
Thanks to Florian R\"ohrer and Matthias Weidenthaler for helping us take data.  We would also like to thank PTB-Berlin for lending us a SQUID-system and lots of help in getting our comagnetometer running.  In particular Jens Voigt for helping us keep the SQUID operational, Wolfgang Kilian for helping us set-up our polarizer and Lutz Trahms and Silvia Knappe-Gr\"{u}neberg for helpful discussions of magnetometry.  We also want to thank Earl Babcock for help on the polarizer and the Lurie Nanofabrication Facility for help producing our cells.  This work was supported by the DFG Cluster of Excellence `Origin and Structure of the Universe' and Michigan State University.  WAT would like to thank the Alexander Von Humboldt foundation for support.  
\end{acknowledgments}

\bibliographystyle{apsrev4-1}
\bibliography{HeXeShifts} 

%merlin.mbs apsrev4-1.bst 2010-07-25 4.21a (PWD, AO, DPC) hacked
%Control: key (0)
%Control: author (72) initials jnrlst
%Control: editor formatted (1) identically to author
%Control: production of article title (-1) disabled
%Control: page (0) single
%Control: year (1) truncated
%Control: production of eprint (0) enabled
\begin{thebibliography}{30}%
\makeatletter
\providecommand \@ifxundefined [1]{%
 \@ifx{#1\undefined}
}%
\providecommand \@ifnum [1]{%
 \ifnum #1\expandafter \@firstoftwo
 \else \expandafter \@secondoftwo
 \fi
}%
\providecommand \@ifx [1]{%
 \ifx #1\expandafter \@firstoftwo
 \else \expandafter \@secondoftwo
 \fi
}%
\providecommand \natexlab [1]{#1}%
\providecommand \enquote  [1]{``#1''}%
\providecommand \bibnamefont  [1]{#1}%
\providecommand \bibfnamefont [1]{#1}%
\providecommand \citenamefont [1]{#1}%
\providecommand \href@noop [0]{\@secondoftwo}%
\providecommand \href [0]{\begingroup \@sanitize@url \@href}%
\providecommand \@href[1]{\@@startlink{#1}\@@href}%
\providecommand \@@href[1]{\endgroup#1\@@endlink}%
\providecommand \@sanitize@url [0]{\catcode `\\12\catcode `\$12\catcode
  `\&12\catcode `\#12\catcode `\^12\catcode `\_12\catcode `\%12\relax}%
\providecommand \@@startlink[1]{}%
\providecommand \@@endlink[0]{}%
\providecommand \url  [0]{\begingroup\@sanitize@url \@url }%
\providecommand \@url [1]{\endgroup\@href {#1}{\urlprefix }}%
\providecommand \urlprefix  [0]{URL }%
\providecommand \Eprint [0]{\href }%
\providecommand \doibase [0]{http://dx.doi.org/}%
\providecommand \selectlanguage [0]{\@gobble}%
\providecommand \bibinfo  [0]{\@secondoftwo}%
\providecommand \bibfield  [0]{\@secondoftwo}%
\providecommand \translation [1]{[#1]}%
\providecommand \BibitemOpen [0]{}%
\providecommand \bibitemStop [0]{}%
\providecommand \bibitemNoStop [0]{.\EOS\space}%
\providecommand \EOS [0]{\spacefactor3000\relax}%
\providecommand \BibitemShut  [1]{\csname bibitem#1\endcsname}%
\let\auto@bib@innerbib\@empty
%</preamble>
\bibitem [{\citenamefont {Chupp}\ \emph {et~al.}(1988)\citenamefont {Chupp},
  \citenamefont {Oteiza}, \citenamefont {Richardson},\ and\ \citenamefont
  {White}}]{Chupp1988}%
  \BibitemOpen
  \bibfield  {author} {\bibinfo {author} {\bibfnamefont {T.~E.}\ \bibnamefont
  {Chupp}}, \bibinfo {author} {\bibfnamefont {E.~R.}\ \bibnamefont {Oteiza}},
  \bibinfo {author} {\bibfnamefont {J.~M.}\ \bibnamefont {Richardson}}, \ and\
  \bibinfo {author} {\bibfnamefont {T.~R.}\ \bibnamefont {White}},\ }\href
  {\doibase 10.1103/PhysRevA.38.3998} {\bibfield  {journal} {\bibinfo
  {journal} {Phys. Rev. A}\ }\textbf {\bibinfo {volume} {38}},\ \bibinfo
  {pages} {3998} (\bibinfo {year} {1988})}\BibitemShut {NoStop}%
\bibitem [{\citenamefont {M\"{o}ller}\ \emph {et~al.}(2002)\citenamefont
  {M\"{o}ller}, \citenamefont {Chen}, \citenamefont {Saam}, \citenamefont
  {Hagspiel}, \citenamefont {Johnson}, \citenamefont {Altes}, \citenamefont
  {de~Lange},\ and\ \citenamefont {Kauczor}}]{Moeller2002}%
  \BibitemOpen
  \bibfield  {author} {\bibinfo {author} {\bibfnamefont {H.~E.}\ \bibnamefont
  {M\"{o}ller}}, \bibinfo {author} {\bibfnamefont {X.~J.}\ \bibnamefont
  {Chen}}, \bibinfo {author} {\bibfnamefont {B.}~\bibnamefont {Saam}}, \bibinfo
  {author} {\bibfnamefont {K.~D.}\ \bibnamefont {Hagspiel}}, \bibinfo {author}
  {\bibfnamefont {G.~A.}\ \bibnamefont {Johnson}}, \bibinfo {author}
  {\bibfnamefont {T.~A.}\ \bibnamefont {Altes}}, \bibinfo {author}
  {\bibfnamefont {E.~E.}\ \bibnamefont {de~Lange}}, \ and\ \bibinfo {author}
  {\bibfnamefont {H.}~\bibnamefont {Kauczor}},\ }\href {\doibase
  10.1002/mrm.10173} {\bibfield  {journal} {\bibinfo  {journal} {Magnetic
  Resonance in Medicine}\ }\textbf {\bibinfo {volume} {47}},\ \bibinfo {pages}
  {1029} (\bibinfo {year} {2002})}\BibitemShut {NoStop}%
\bibitem [{\citenamefont {Meersmann}\ and\ \citenamefont
  {Brunner}(2015)}]{Meersman2015}%
  \BibitemOpen
  \bibinfo {editor} {\bibfnamefont {T.}~\bibnamefont {Meersmann}}\ and\
  \bibinfo {editor} {\bibfnamefont {E.}~\bibnamefont {Brunner}},\ eds.,\ \href
  {\doibase 10.1039/9781782628378} {\emph {\bibinfo {title} {Hyperpolarized
  Xenon-129 Magnetic Resonance}}},\ New Developments in NMR\ (\bibinfo
  {publisher} {The Royal Society of Chemistry},\ \bibinfo {year} {2015})\ pp.\
  \bibinfo {pages} {P001--484}\BibitemShut {NoStop}%
\bibitem [{\citenamefont {Kornack}\ \emph {et~al.}(2005)\citenamefont
  {Kornack}, \citenamefont {Ghosh},\ and\ \citenamefont
  {Romalis}}]{Kornack2005}%
  \BibitemOpen
  \bibfield  {author} {\bibinfo {author} {\bibfnamefont {T.~W.}\ \bibnamefont
  {Kornack}}, \bibinfo {author} {\bibfnamefont {R.~K.}\ \bibnamefont {Ghosh}},
  \ and\ \bibinfo {author} {\bibfnamefont {M.~V.}\ \bibnamefont {Romalis}},\
  }\href {\doibase 10.1103/PhysRevLett.95.230801} {\bibfield  {journal}
  {\bibinfo  {journal} {Phys. Rev. Lett.}\ }\textbf {\bibinfo {volume} {95}},\
  \bibinfo {pages} {230801} (\bibinfo {year} {2005})}\BibitemShut {NoStop}%
\bibitem [{\citenamefont {{Vasilakis}}\ \emph {et~al.}(2009)\citenamefont
  {{Vasilakis}}, \citenamefont {{Brown}}, \citenamefont {{Kornack}},\ and\
  \citenamefont {{Romalis}}}]{Vasilakis2009}%
  \BibitemOpen
  \bibfield  {author} {\bibinfo {author} {\bibfnamefont {G.}~\bibnamefont
  {{Vasilakis}}}, \bibinfo {author} {\bibfnamefont {J.~M.}\ \bibnamefont
  {{Brown}}}, \bibinfo {author} {\bibfnamefont {T.~W.}\ \bibnamefont
  {{Kornack}}}, \ and\ \bibinfo {author} {\bibfnamefont {M.~V.}\ \bibnamefont
  {{Romalis}}},\ }\href {\doibase 10.1103/PhysRevLett.103.261801} {\bibfield
  {journal} {\bibinfo  {journal} {Physical Review Letters}\ }\textbf {\bibinfo
  {volume} {103}},\ \bibinfo {eid} {261801} (\bibinfo {year} {2009})},\ \Eprint
  {http://arxiv.org/abs/0809.4700} {arXiv:0809.4700 [physics.atom-ph]}
  \BibitemShut {NoStop}%
\bibitem [{\citenamefont {{Brown}}\ \emph {et~al.}(2010)\citenamefont
  {{Brown}}, \citenamefont {{Smullin}}, \citenamefont {{Kornack}},\ and\
  \citenamefont {{Romalis}}}]{Brown2010}%
  \BibitemOpen
  \bibfield  {author} {\bibinfo {author} {\bibfnamefont {J.~M.}\ \bibnamefont
  {{Brown}}}, \bibinfo {author} {\bibfnamefont {S.~J.}\ \bibnamefont
  {{Smullin}}}, \bibinfo {author} {\bibfnamefont {T.~W.}\ \bibnamefont
  {{Kornack}}}, \ and\ \bibinfo {author} {\bibfnamefont {M.~V.}\ \bibnamefont
  {{Romalis}}},\ }\href {\doibase 10.1103/PhysRevLett.105.151604} {\bibfield
  {journal} {\bibinfo  {journal} {Physical Review Letters}\ }\textbf {\bibinfo
  {volume} {105}},\ \bibinfo {eid} {151604} (\bibinfo {year} {2010})},\ \Eprint
  {http://arxiv.org/abs/1006.5425} {arXiv:1006.5425 [physics.atom-ph]}
  \BibitemShut {NoStop}%
\bibitem [{\citenamefont {{Smiciklas}}\ \emph {et~al.}(2011)\citenamefont
  {{Smiciklas}}, \citenamefont {{Brown}}, \citenamefont {{Cheuk}},
  \citenamefont {{Smullin}},\ and\ \citenamefont {{Romalis}}}]{Smiciklas2011}%
  \BibitemOpen
  \bibfield  {author} {\bibinfo {author} {\bibfnamefont {M.}~\bibnamefont
  {{Smiciklas}}}, \bibinfo {author} {\bibfnamefont {J.~M.}\ \bibnamefont
  {{Brown}}}, \bibinfo {author} {\bibfnamefont {L.~W.}\ \bibnamefont
  {{Cheuk}}}, \bibinfo {author} {\bibfnamefont {S.~J.}\ \bibnamefont
  {{Smullin}}}, \ and\ \bibinfo {author} {\bibfnamefont {M.~V.}\ \bibnamefont
  {{Romalis}}},\ }\href {\doibase 10.1103/PhysRevLett.107.171604} {\bibfield
  {journal} {\bibinfo  {journal} {Physical Review Letters}\ }\textbf {\bibinfo
  {volume} {107}},\ \bibinfo {eid} {171604} (\bibinfo {year} {2011})},\ \Eprint
  {http://arxiv.org/abs/1106.0738} {arXiv:1106.0738 [physics.atom-ph]}
  \BibitemShut {NoStop}%
\bibitem [{\citenamefont {Lee}\ \emph {et~al.}(2018)\citenamefont {Lee},
  \citenamefont {Almasi},\ and\ \citenamefont {Romalis}}]{Lee2018}%
  \BibitemOpen
  \bibfield  {author} {\bibinfo {author} {\bibfnamefont {J.}~\bibnamefont
  {Lee}}, \bibinfo {author} {\bibfnamefont {A.}~\bibnamefont {Almasi}}, \ and\
  \bibinfo {author} {\bibfnamefont {M.}~\bibnamefont {Romalis}},\ }\href
  {\doibase 10.1103/PhysRevLett.120.161801} {\bibfield  {journal} {\bibinfo
  {journal} {Phys. Rev. Lett.}\ }\textbf {\bibinfo {volume} {120}},\ \bibinfo
  {pages} {161801} (\bibinfo {year} {2018})}\BibitemShut {NoStop}%
\bibitem [{\citenamefont {Bear}\ \emph {et~al.}(2000)\citenamefont {Bear},
  \citenamefont {Stoner}, \citenamefont {Walsworth}, \citenamefont
  {Kosteleck\'y},\ and\ \citenamefont {Lane}}]{Bear2000}%
  \BibitemOpen
  \bibfield  {author} {\bibinfo {author} {\bibfnamefont {D.}~\bibnamefont
  {Bear}}, \bibinfo {author} {\bibfnamefont {R.~E.}\ \bibnamefont {Stoner}},
  \bibinfo {author} {\bibfnamefont {R.~L.}\ \bibnamefont {Walsworth}}, \bibinfo
  {author} {\bibfnamefont {V.~A.}\ \bibnamefont {Kosteleck\'y}}, \ and\
  \bibinfo {author} {\bibfnamefont {C.~D.}\ \bibnamefont {Lane}},\ }\href
  {\doibase 10.1103/PhysRevLett.85.5038} {\bibfield  {journal} {\bibinfo
  {journal} {Phys. Rev. Lett.}\ }\textbf {\bibinfo {volume} {85}},\ \bibinfo
  {pages} {5038} (\bibinfo {year} {2000})}\BibitemShut {NoStop}%
\bibitem [{\citenamefont {Gemmel}\ \emph
  {et~al.}(2010{\natexlab{a}})\citenamefont {Gemmel}, \citenamefont {Heil},
  \citenamefont {Karpuk}, \citenamefont {Lenz}, \citenamefont {Sobolev},
  \citenamefont {Tullney}, \citenamefont {Burghoff}, \citenamefont {Kilian},
  \citenamefont {Knappe-Gr\"uneberg}, \citenamefont {M\"uller}, \citenamefont
  {Schnabel}, \citenamefont {Seifert}, \citenamefont {Trahms},\ and\
  \citenamefont {Schmidt}}]{Gemmel2010}%
  \BibitemOpen
  \bibfield  {author} {\bibinfo {author} {\bibfnamefont {C.}~\bibnamefont
  {Gemmel}}, \bibinfo {author} {\bibfnamefont {W.}~\bibnamefont {Heil}},
  \bibinfo {author} {\bibfnamefont {S.}~\bibnamefont {Karpuk}}, \bibinfo
  {author} {\bibfnamefont {K.}~\bibnamefont {Lenz}}, \bibinfo {author}
  {\bibfnamefont {Y.}~\bibnamefont {Sobolev}}, \bibinfo {author} {\bibfnamefont
  {K.}~\bibnamefont {Tullney}}, \bibinfo {author} {\bibfnamefont
  {M.}~\bibnamefont {Burghoff}}, \bibinfo {author} {\bibfnamefont
  {W.}~\bibnamefont {Kilian}}, \bibinfo {author} {\bibfnamefont
  {S.}~\bibnamefont {Knappe-Gr\"uneberg}}, \bibinfo {author} {\bibfnamefont
  {W.}~\bibnamefont {M\"uller}}, \bibinfo {author} {\bibfnamefont
  {A.}~\bibnamefont {Schnabel}}, \bibinfo {author} {\bibfnamefont
  {F.}~\bibnamefont {Seifert}}, \bibinfo {author} {\bibfnamefont
  {L.}~\bibnamefont {Trahms}}, \ and\ \bibinfo {author} {\bibfnamefont
  {U.}~\bibnamefont {Schmidt}},\ }\href {\doibase 10.1103/PhysRevD.82.111901}
  {\bibfield  {journal} {\bibinfo  {journal} {Phys. Rev. D}\ }\textbf {\bibinfo
  {volume} {82}},\ \bibinfo {pages} {111901} (\bibinfo {year}
  {2010}{\natexlab{a}})}\BibitemShut {NoStop}%
\bibitem [{\citenamefont {Allmendinger}\ \emph
  {et~al.}(2014{\natexlab{a}})\citenamefont {Allmendinger}, \citenamefont
  {Heil}, \citenamefont {Karpuk}, \citenamefont {Kilian}, \citenamefont
  {Scharth}, \citenamefont {Schmidt}, \citenamefont {Schnabel}, \citenamefont
  {Sobolev},\ and\ \citenamefont {Tullney}}]{Allmendinger2014}%
  \BibitemOpen
  \bibfield  {author} {\bibinfo {author} {\bibfnamefont {F.}~\bibnamefont
  {Allmendinger}}, \bibinfo {author} {\bibfnamefont {W.}~\bibnamefont {Heil}},
  \bibinfo {author} {\bibfnamefont {S.}~\bibnamefont {Karpuk}}, \bibinfo
  {author} {\bibfnamefont {W.}~\bibnamefont {Kilian}}, \bibinfo {author}
  {\bibfnamefont {A.}~\bibnamefont {Scharth}}, \bibinfo {author} {\bibfnamefont
  {U.}~\bibnamefont {Schmidt}}, \bibinfo {author} {\bibfnamefont
  {A.}~\bibnamefont {Schnabel}}, \bibinfo {author} {\bibfnamefont
  {Y.}~\bibnamefont {Sobolev}}, \ and\ \bibinfo {author} {\bibfnamefont
  {K.}~\bibnamefont {Tullney}},\ }\href {\doibase
  10.1103/PhysRevLett.112.110801} {\bibfield  {journal} {\bibinfo  {journal}
  {Phys. Rev. Lett.}\ }\textbf {\bibinfo {volume} {112}},\ \bibinfo {pages}
  {110801} (\bibinfo {year} {2014}{\natexlab{a}})}\BibitemShut {NoStop}%
\bibitem [{\citenamefont {Rosenberry}\ and\ \citenamefont
  {Chupp}(2001)}]{Rosenberry2001}%
  \BibitemOpen
  \bibfield  {author} {\bibinfo {author} {\bibfnamefont {M.~A.}\ \bibnamefont
  {Rosenberry}}\ and\ \bibinfo {author} {\bibfnamefont {T.~E.}\ \bibnamefont
  {Chupp}},\ }\href {\doibase 10.1103/PhysRevLett.86.22} {\bibfield  {journal}
  {\bibinfo  {journal} {Phys. Rev. Lett.}\ }\textbf {\bibinfo {volume} {86}},\
  \bibinfo {pages} {22} (\bibinfo {year} {2001})}\BibitemShut {NoStop}%
\bibitem [{\citenamefont {{Tullney}}\ \emph {et~al.}(2013)\citenamefont
  {{Tullney}}, \citenamefont {{Allmendinger}}, \citenamefont {{Burghoff}},
  \citenamefont {{Heil}}, \citenamefont {{Karpuk}}, \citenamefont {{Kilian}},
  \citenamefont {{Knappe-Gr{\"u}neberg}}, \citenamefont {{M{\"u}ller}},
  \citenamefont {{Schmidt}}, \citenamefont {{Schnabel}}, \citenamefont
  {{Seifert}}, \citenamefont {{Sobolev}},\ and\ \citenamefont
  {{Trahms}}}]{Tullney2013}%
  \BibitemOpen
  \bibfield  {author} {\bibinfo {author} {\bibfnamefont {K.}~\bibnamefont
  {{Tullney}}}, \bibinfo {author} {\bibfnamefont {F.}~\bibnamefont
  {{Allmendinger}}}, \bibinfo {author} {\bibfnamefont {M.}~\bibnamefont
  {{Burghoff}}}, \bibinfo {author} {\bibfnamefont {W.}~\bibnamefont {{Heil}}},
  \bibinfo {author} {\bibfnamefont {S.}~\bibnamefont {{Karpuk}}}, \bibinfo
  {author} {\bibfnamefont {W.}~\bibnamefont {{Kilian}}}, \bibinfo {author}
  {\bibfnamefont {S.}~\bibnamefont {{Knappe-Gr{\"u}neberg}}}, \bibinfo {author}
  {\bibfnamefont {W.}~\bibnamefont {{M{\"u}ller}}}, \bibinfo {author}
  {\bibfnamefont {U.}~\bibnamefont {{Schmidt}}}, \bibinfo {author}
  {\bibfnamefont {A.}~\bibnamefont {{Schnabel}}}, \bibinfo {author}
  {\bibfnamefont {F.}~\bibnamefont {{Seifert}}}, \bibinfo {author}
  {\bibfnamefont {Y.}~\bibnamefont {{Sobolev}}}, \ and\ \bibinfo {author}
  {\bibfnamefont {L.}~\bibnamefont {{Trahms}}},\ }\href {\doibase
  10.1103/PhysRevLett.111.100801} {\bibfield  {journal} {\bibinfo  {journal}
  {Physical Review Letters}\ }\textbf {\bibinfo {volume} {111}},\ \bibinfo
  {eid} {100801} (\bibinfo {year} {2013})},\ \Eprint
  {http://arxiv.org/abs/1303.6612} {arXiv:1303.6612 [hep-ex]} \BibitemShut
  {NoStop}%
\bibitem [{\citenamefont {Graham}\ \emph {et~al.}(2018)\citenamefont {Graham},
  \citenamefont {Kaplan}, \citenamefont {Mardon}, \citenamefont {Rajendran},
  \citenamefont {Terrano}, \citenamefont {Trahms},\ and\ \citenamefont
  {Wilkason}}]{Graham2018}%
  \BibitemOpen
  \bibfield  {author} {\bibinfo {author} {\bibfnamefont {P.~W.}\ \bibnamefont
  {Graham}}, \bibinfo {author} {\bibfnamefont {D.~E.}\ \bibnamefont {Kaplan}},
  \bibinfo {author} {\bibfnamefont {J.}~\bibnamefont {Mardon}}, \bibinfo
  {author} {\bibfnamefont {S.}~\bibnamefont {Rajendran}}, \bibinfo {author}
  {\bibfnamefont {W.~A.}\ \bibnamefont {Terrano}}, \bibinfo {author}
  {\bibfnamefont {L.}~\bibnamefont {Trahms}}, \ and\ \bibinfo {author}
  {\bibfnamefont {T.}~\bibnamefont {Wilkason}},\ }\href {\doibase
  10.1103/PhysRevD.97.055006} {\bibfield  {journal} {\bibinfo  {journal} {Phys.
  Rev. D}\ }\textbf {\bibinfo {volume} {97}},\ \bibinfo {pages} {055006}
  (\bibinfo {year} {2018})}\BibitemShut {NoStop}%
\bibitem [{\citenamefont {{Limes}}\ \emph {et~al.}(2018)\citenamefont
  {{Limes}}, \citenamefont {{Dural}}, \citenamefont {{Romalis}}, \citenamefont
  {{Foley}}, \citenamefont {{Kornack}}, \citenamefont {{Nelson}},\ and\
  \citenamefont {{Grisham}}}]{Limes2018b}%
  \BibitemOpen
  \bibfield  {author} {\bibinfo {author} {\bibfnamefont {M.~E.}\ \bibnamefont
  {{Limes}}}, \bibinfo {author} {\bibfnamefont {N.}~\bibnamefont {{Dural}}},
  \bibinfo {author} {\bibfnamefont {M.~V.}\ \bibnamefont {{Romalis}}}, \bibinfo
  {author} {\bibfnamefont {E.~L.}\ \bibnamefont {{Foley}}}, \bibinfo {author}
  {\bibfnamefont {T.~W.}\ \bibnamefont {{Kornack}}}, \bibinfo {author}
  {\bibfnamefont {A.}~\bibnamefont {{Nelson}}}, \ and\ \bibinfo {author}
  {\bibfnamefont {L.~R.}\ \bibnamefont {{Grisham}}},\ }\href@noop {} {\bibfield
   {journal} {\bibinfo  {journal} {ArXiv e-prints}\ } (\bibinfo {year}
  {2018})},\ \Eprint {http://arxiv.org/abs/1805.11578} {arXiv:1805.11578
  [physics.atom-ph]} \BibitemShut {NoStop}%
\bibitem [{\citenamefont {Koch}\ \emph {et~al.}(2015)\citenamefont {Koch},
  \citenamefont {Bison}, \citenamefont {Gruji{\'{c}}}, \citenamefont {Heil},
  \citenamefont {Kasprzak}, \citenamefont {Knowles}, \citenamefont {Kraft},
  \citenamefont {Pazgalev}, \citenamefont {Schnabel}, \citenamefont {Voigt},\
  and\ \citenamefont {Weis}}]{Koch2015}%
  \BibitemOpen
  \bibfield  {author} {\bibinfo {author} {\bibfnamefont {H.}~\bibnamefont
  {Koch}}, \bibinfo {author} {\bibfnamefont {G.}~\bibnamefont {Bison}},
  \bibinfo {author} {\bibfnamefont {Z.~D.}\ \bibnamefont {Gruji{\'{c}}}},
  \bibinfo {author} {\bibfnamefont {W.}~\bibnamefont {Heil}}, \bibinfo {author}
  {\bibfnamefont {M.}~\bibnamefont {Kasprzak}}, \bibinfo {author}
  {\bibfnamefont {P.}~\bibnamefont {Knowles}}, \bibinfo {author} {\bibfnamefont
  {A.}~\bibnamefont {Kraft}}, \bibinfo {author} {\bibfnamefont
  {A.}~\bibnamefont {Pazgalev}}, \bibinfo {author} {\bibfnamefont
  {A.}~\bibnamefont {Schnabel}}, \bibinfo {author} {\bibfnamefont
  {J.}~\bibnamefont {Voigt}}, \ and\ \bibinfo {author} {\bibfnamefont
  {A.}~\bibnamefont {Weis}},\ }\href {\doibase 10.1140/epjd/e2015-60018-7}
  {\bibfield  {journal} {\bibinfo  {journal} {The European Physical Journal D}\
  }\textbf {\bibinfo {volume} {69}},\ \bibinfo {pages} {202} (\bibinfo {year}
  {2015})}\BibitemShut {NoStop}%
\bibitem [{\citenamefont {{Katz}}\ and\ \citenamefont
  {{Firstenberg}}(2018)}]{Katz2018}%
  \BibitemOpen
  \bibfield  {author} {\bibinfo {author} {\bibfnamefont {O.}~\bibnamefont
  {{Katz}}}\ and\ \bibinfo {author} {\bibfnamefont {O.}~\bibnamefont
  {{Firstenberg}}},\ }\href {\doibase 10.1038/s41467-018-04458-4} {\bibfield
  {journal} {\bibinfo  {journal} {Nature Communications}\ }\textbf {\bibinfo
  {volume} {9}},\ \bibinfo {eid} {2074} (\bibinfo {year} {2018})},\ \Eprint
  {http://arxiv.org/abs/1710.06844} {arXiv:1710.06844 [quant-ph]} \BibitemShut
  {NoStop}%
\bibitem [{\citenamefont {Romalis}\ \emph {et~al.}(2014)\citenamefont
  {Romalis}, \citenamefont {Sheng}, \citenamefont {Saam},\ and\ \citenamefont
  {Walker}}]{Romalis2014}%
  \BibitemOpen
  \bibfield  {author} {\bibinfo {author} {\bibfnamefont {M.~V.}\ \bibnamefont
  {Romalis}}, \bibinfo {author} {\bibfnamefont {D.}~\bibnamefont {Sheng}},
  \bibinfo {author} {\bibfnamefont {B.}~\bibnamefont {Saam}}, \ and\ \bibinfo
  {author} {\bibfnamefont {T.~G.}\ \bibnamefont {Walker}},\ }\href {\doibase
  10.1103/PhysRevLett.113.188901} {\bibfield  {journal} {\bibinfo  {journal}
  {Phys. Rev. Lett.}\ }\textbf {\bibinfo {volume} {113}},\ \bibinfo {pages}
  {188901} (\bibinfo {year} {2014})}\BibitemShut {NoStop}%
\bibitem [{\citenamefont {Allmendinger}\ \emph
  {et~al.}(2014{\natexlab{b}})\citenamefont {Allmendinger}, \citenamefont
  {Schmidt}, \citenamefont {Heil}, \citenamefont {Karpuk}, \citenamefont
  {Scharth}, \citenamefont {Sobolev},\ and\ \citenamefont
  {Tullney}}]{Allmendinger2014b}%
  \BibitemOpen
  \bibfield  {author} {\bibinfo {author} {\bibfnamefont {F.}~\bibnamefont
  {Allmendinger}}, \bibinfo {author} {\bibfnamefont {U.}~\bibnamefont
  {Schmidt}}, \bibinfo {author} {\bibfnamefont {W.}~\bibnamefont {Heil}},
  \bibinfo {author} {\bibfnamefont {S.}~\bibnamefont {Karpuk}}, \bibinfo
  {author} {\bibfnamefont {A.}~\bibnamefont {Scharth}}, \bibinfo {author}
  {\bibfnamefont {Y.}~\bibnamefont {Sobolev}}, \ and\ \bibinfo {author}
  {\bibfnamefont {K.}~\bibnamefont {Tullney}},\ }\href {\doibase
  10.1103/PhysRevLett.113.188902} {\bibfield  {journal} {\bibinfo  {journal}
  {Phys. Rev. Lett.}\ }\textbf {\bibinfo {volume} {113}},\ \bibinfo {pages}
  {188902} (\bibinfo {year} {2014}{\natexlab{b}})}\BibitemShut {NoStop}%
\bibitem [{\citenamefont {Gemmel}\ \emph
  {et~al.}(2010{\natexlab{b}})\citenamefont {Gemmel}, \citenamefont {Heil},
  \citenamefont {Karpuk}, \citenamefont {Lenz}, \citenamefont {Ludwig},
  \citenamefont {Sobolev}, \citenamefont {Tullney}, \citenamefont {Burghoff},
  \citenamefont {Kilian}, \citenamefont {Knappe-Gr{\"u}neberg}, \citenamefont
  {M{\"u}ller}, \citenamefont {Schnabel}, \citenamefont {Seifert},
  \citenamefont {Trahms},\ and\ \citenamefont {Bae{\ss}ler}}]{Gemmel2010b}%
  \BibitemOpen
  \bibfield  {author} {\bibinfo {author} {\bibfnamefont {C.}~\bibnamefont
  {Gemmel}}, \bibinfo {author} {\bibfnamefont {W.}~\bibnamefont {Heil}},
  \bibinfo {author} {\bibfnamefont {S.}~\bibnamefont {Karpuk}}, \bibinfo
  {author} {\bibfnamefont {K.}~\bibnamefont {Lenz}}, \bibinfo {author}
  {\bibfnamefont {C.}~\bibnamefont {Ludwig}}, \bibinfo {author} {\bibfnamefont
  {Y.}~\bibnamefont {Sobolev}}, \bibinfo {author} {\bibfnamefont
  {K.}~\bibnamefont {Tullney}}, \bibinfo {author} {\bibfnamefont
  {M.}~\bibnamefont {Burghoff}}, \bibinfo {author} {\bibfnamefont
  {W.}~\bibnamefont {Kilian}}, \bibinfo {author} {\bibfnamefont
  {S.}~\bibnamefont {Knappe-Gr{\"u}neberg}}, \bibinfo {author} {\bibfnamefont
  {W.}~\bibnamefont {M{\"u}ller}}, \bibinfo {author} {\bibfnamefont
  {A.}~\bibnamefont {Schnabel}}, \bibinfo {author} {\bibfnamefont
  {F.}~\bibnamefont {Seifert}}, \bibinfo {author} {\bibfnamefont
  {L.}~\bibnamefont {Trahms}}, \ and\ \bibinfo {author} {\bibfnamefont
  {S.}~\bibnamefont {Bae{\ss}ler}},\ }\href {\doibase
  10.1140/epjd/e2010-00044-5} {\bibfield  {journal} {\bibinfo  {journal} {The
  European Physical Journal D}\ }\textbf {\bibinfo {volume} {57}},\ \bibinfo
  {pages} {303} (\bibinfo {year} {2010}{\natexlab{b}})}\BibitemShut {NoStop}%
\bibitem [{\citenamefont {Rosenberry}(2000)}]{Rosenberry2000}%
  \BibitemOpen
  \bibfield  {author} {\bibinfo {author} {\bibfnamefont {M.~A.}\ \bibnamefont
  {Rosenberry}},\ }\emph {\bibinfo {title} {A Precision Measurement of the
  $^{129}$Xe Electric Dipole Moment Using Dual Noble Gas Masers}},\ \href@noop
  {} {Ph.D. thesis},\ \bibinfo  {school} {University of Michigan} (\bibinfo
  {year} {2000})\BibitemShut {NoStop}%
\bibitem [{\citenamefont {Oteiza}(1992)}]{Oteiza1992}%
  \BibitemOpen
  \bibfield  {author} {\bibinfo {author} {\bibfnamefont {E.~R.}\ \bibnamefont
  {Oteiza}},\ }\emph {\bibinfo {title} {Search for a Permanent Electric Dipole
  Moment in $^{129}$Xe Using Simultaneous $^3$He Magnetometry}},\ \href@noop {}
  {Ph.D. thesis},\ \bibinfo  {school} {Harvard University} (\bibinfo {year}
  {1992})\BibitemShut {NoStop}%
\bibitem [{\citenamefont {Schaefer}\ \emph {et~al.}(1989)\citenamefont
  {Schaefer}, \citenamefont {Cates}, \citenamefont {Chien}, \citenamefont
  {Gonatas}, \citenamefont {Happer},\ and\ \citenamefont
  {Walker}}]{Schaefer1989kappa}%
  \BibitemOpen
  \bibfield  {author} {\bibinfo {author} {\bibfnamefont {S.~R.}\ \bibnamefont
  {Schaefer}}, \bibinfo {author} {\bibfnamefont {G.~D.}\ \bibnamefont {Cates}},
  \bibinfo {author} {\bibfnamefont {T.-R.}\ \bibnamefont {Chien}}, \bibinfo
  {author} {\bibfnamefont {D.}~\bibnamefont {Gonatas}}, \bibinfo {author}
  {\bibfnamefont {W.}~\bibnamefont {Happer}}, \ and\ \bibinfo {author}
  {\bibfnamefont {T.~G.}\ \bibnamefont {Walker}},\ }\href {\doibase
  10.1103/PhysRevA.39.5613} {\bibfield  {journal} {\bibinfo  {journal} {Phys.
  Rev. A}\ }\textbf {\bibinfo {volume} {39}},\ \bibinfo {pages} {5613}
  (\bibinfo {year} {1989})}\BibitemShut {NoStop}%
\bibitem [{\citenamefont {Vlassenbroek}\ \emph {et~al.}(1996)\citenamefont
  {Vlassenbroek}, \citenamefont {Jeener},\ and\ \citenamefont
  {Broekaert}}]{Vlassenbroek1996}%
  \BibitemOpen
  \bibfield  {author} {\bibinfo {author} {\bibfnamefont {A.}~\bibnamefont
  {Vlassenbroek}}, \bibinfo {author} {\bibfnamefont {J.}~\bibnamefont
  {Jeener}}, \ and\ \bibinfo {author} {\bibfnamefont {P.}~\bibnamefont
  {Broekaert}},\ }\href {\doibase https://doi.org/10.1006/jmra.1996.0032}
  {\bibfield  {journal} {\bibinfo  {journal} {Journal of Magnetic Resonance,
  Series A}\ }\textbf {\bibinfo {volume} {118}},\ \bibinfo {pages} {234 }
  (\bibinfo {year} {1996})}\BibitemShut {NoStop}%
\bibitem [{\citenamefont {Walker}\ and\ \citenamefont
  {Happer}(1997)}]{Walker1997}%
  \BibitemOpen
  \bibfield  {author} {\bibinfo {author} {\bibfnamefont {T.~G.}\ \bibnamefont
  {Walker}}\ and\ \bibinfo {author} {\bibfnamefont {W.}~\bibnamefont
  {Happer}},\ }\href {\doibase 10.1103/RevModPhys.69.629} {\bibfield  {journal}
  {\bibinfo  {journal} {Rev. Mod. Phys.}\ }\textbf {\bibinfo {volume} {69}},\
  \bibinfo {pages} {629} (\bibinfo {year} {1997})}\BibitemShut {NoStop}%
\bibitem [{\citenamefont {Gentile}\ \emph {et~al.}(2017)\citenamefont
  {Gentile}, \citenamefont {Nacher}, \citenamefont {Saam},\ and\ \citenamefont
  {Walker}}]{Gentile2017}%
  \BibitemOpen
  \bibfield  {author} {\bibinfo {author} {\bibfnamefont {T.~R.}\ \bibnamefont
  {Gentile}}, \bibinfo {author} {\bibfnamefont {P.~J.}\ \bibnamefont {Nacher}},
  \bibinfo {author} {\bibfnamefont {B.}~\bibnamefont {Saam}}, \ and\ \bibinfo
  {author} {\bibfnamefont {T.~G.}\ \bibnamefont {Walker}},\ }\href {\doibase
  10.1103/RevModPhys.89.045004} {\bibfield  {journal} {\bibinfo  {journal}
  {Rev. Mod. Phys.}\ }\textbf {\bibinfo {volume} {89}},\ \bibinfo {pages}
  {045004} (\bibinfo {year} {2017})}\BibitemShut {NoStop}%
\bibitem [{\citenamefont {Kuchler}\ \emph {et~al.}(2016)\citenamefont
  {Kuchler}, \citenamefont {Babcock}, \citenamefont {Burghoff}, \citenamefont
  {Chupp}, \citenamefont {Degenkolb}, \citenamefont {Fan}, \citenamefont
  {Fierlinger}, \citenamefont {Gong}, \citenamefont {Kraegeloh}, \citenamefont
  {Kilian}, \citenamefont {Knappe-Gr{\"u}neberg}, \citenamefont {Lins},
  \citenamefont {Marino}, \citenamefont {Meinel}, \citenamefont {Niessen},
  \citenamefont {Sachdeva}, \citenamefont {Salhi}, \citenamefont {Schnabel},
  \citenamefont {Seifert}, \citenamefont {Singh}, \citenamefont {Stuiber},
  \citenamefont {Trahms},\ and\ \citenamefont {Voigt}}]{Kuchler2016}%
  \BibitemOpen
  \bibfield  {author} {\bibinfo {author} {\bibfnamefont {F.}~\bibnamefont
  {Kuchler}}, \bibinfo {author} {\bibfnamefont {E.}~\bibnamefont {Babcock}},
  \bibinfo {author} {\bibfnamefont {M.}~\bibnamefont {Burghoff}}, \bibinfo
  {author} {\bibfnamefont {T.}~\bibnamefont {Chupp}}, \bibinfo {author}
  {\bibfnamefont {S.}~\bibnamefont {Degenkolb}}, \bibinfo {author}
  {\bibfnamefont {I.}~\bibnamefont {Fan}}, \bibinfo {author} {\bibfnamefont
  {P.}~\bibnamefont {Fierlinger}}, \bibinfo {author} {\bibfnamefont
  {F.}~\bibnamefont {Gong}}, \bibinfo {author} {\bibfnamefont {E.}~\bibnamefont
  {Kraegeloh}}, \bibinfo {author} {\bibfnamefont {W.}~\bibnamefont {Kilian}},
  \bibinfo {author} {\bibfnamefont {S.}~\bibnamefont {Knappe-Gr{\"u}neberg}},
  \bibinfo {author} {\bibfnamefont {T.}~\bibnamefont {Lins}}, \bibinfo {author}
  {\bibfnamefont {M.}~\bibnamefont {Marino}}, \bibinfo {author} {\bibfnamefont
  {J.}~\bibnamefont {Meinel}}, \bibinfo {author} {\bibfnamefont
  {B.}~\bibnamefont {Niessen}}, \bibinfo {author} {\bibfnamefont
  {N.}~\bibnamefont {Sachdeva}}, \bibinfo {author} {\bibfnamefont
  {Z.}~\bibnamefont {Salhi}}, \bibinfo {author} {\bibfnamefont
  {A.}~\bibnamefont {Schnabel}}, \bibinfo {author} {\bibfnamefont
  {F.}~\bibnamefont {Seifert}}, \bibinfo {author} {\bibfnamefont
  {J.}~\bibnamefont {Singh}}, \bibinfo {author} {\bibfnamefont
  {S.}~\bibnamefont {Stuiber}}, \bibinfo {author} {\bibfnamefont
  {L.}~\bibnamefont {Trahms}}, \ and\ \bibinfo {author} {\bibfnamefont
  {J.}~\bibnamefont {Voigt}},\ }\href {\doibase 10.1007/s10751-016-1302-9}
  {\bibfield  {journal} {\bibinfo  {journal} {Hyperfine Interactions}\ }\textbf
  {\bibinfo {volume} {237}},\ \bibinfo {pages} {95} (\bibinfo {year}
  {2016})}\BibitemShut {NoStop}%
\bibitem [{\citenamefont {Drung}(2002)}]{Drung2002}%
  \BibitemOpen
  \bibfield  {author} {\bibinfo {author} {\bibfnamefont {D.}~\bibnamefont
  {Drung}},\ }\href {\doibase https://doi.org/10.1016/S0921-4534(01)01154-6}
  {\bibfield  {journal} {\bibinfo  {journal} {Physica C: Superconductivity}\
  }\textbf {\bibinfo {volume} {368}},\ \bibinfo {pages} {134 } (\bibinfo {year}
  {2002})}\BibitemShut {NoStop}%
\bibitem [{\citenamefont {Sheng}\ \emph {et~al.}(2014)\citenamefont {Sheng},
  \citenamefont {Kabcenell},\ and\ \citenamefont {Romalis}}]{Sheng2014}%
  \BibitemOpen
  \bibfield  {author} {\bibinfo {author} {\bibfnamefont {D.}~\bibnamefont
  {Sheng}}, \bibinfo {author} {\bibfnamefont {A.}~\bibnamefont {Kabcenell}}, \
  and\ \bibinfo {author} {\bibfnamefont {M.~V.}\ \bibnamefont {Romalis}},\
  }\href {\doibase 10.1103/PhysRevLett.113.163002} {\bibfield  {journal}
  {\bibinfo  {journal} {Phys. Rev. Lett.}\ }\textbf {\bibinfo {volume} {113}},\
  \bibinfo {pages} {163002} (\bibinfo {year} {2014})}\BibitemShut {NoStop}%
\bibitem [{\citenamefont {{Vaara}}\ and\ \citenamefont
  {{Romalis}}(2018)}]{Vaara2018}%
  \BibitemOpen
  \bibfield  {author} {\bibinfo {author} {\bibfnamefont {J.}~\bibnamefont
  {{Vaara}}}\ and\ \bibinfo {author} {\bibfnamefont {M.~V.}\ \bibnamefont
  {{Romalis}}},\ }\href@noop {} {\bibfield  {journal} {\bibinfo  {journal}
  {ArXiv e-prints}\ } (\bibinfo {year} {2018})},\ \Eprint
  {http://arxiv.org/abs/1811.08678} {arXiv:1811.08678 [physics.atom-ph]}
  \BibitemShut {NoStop}%
\end{thebibliography}%

\setlength{\extrarowheight}{5pt}
\begin{table*}[h]
\caption{Results from our study of species-specific, longitudinal-magnetization-dependent frequency shifts.
\emph{Ratio} gives the species for which we measured $\Delta^{\!(m)} \mathcal{R}_k = \Delta^m (\widetilde{\omega}/\omega_k)$: the change in the corrected to absolute frequency ratio of species $k$ when the longitudinal magnetization $M^\mf{L}_m$ of species $m$ is flipped.  $\alpha$ is the angle of the magnetic field to the cell axis.  Model is the theoretical expectation assuming only a scalar $\kappa$ interaction and resonant geometric effect (Eq. \ref{eq: error}), with $\Gamma_0$ defined by $B_\mf{int}^\mf{L}(\alpha) = \mu_0 M^\mf{L} \Gamma_0 (3 \cos^2 \alpha - 1)$ and measured to be $\Gamma_0 = 0.023\pm0.002$ from the absolute magnitudes of the homonuclear shifts.  Measured value is the value of $\Delta^{\!(m)} \mathcal{R}_k$ from our full data set. Extracted $\kappa$ is the $\kappa$ consistent with $\Delta^{\!(m)} \mathcal{R}_k$ assuming the model. Fills is the number of separate cell fillings of different pressures that contributed to the measured value, and Shifts is the number of independent measurements of the $\Delta^{\!(m)} \mathcal{R}_k$ that we made by flipping $M^\mf{L}_m$.}
\begin{center}
\begin{tabular}{ccccccc}
Ratio & $\alpha$ & Model & Measured value & Extracted $\kappa$ & Fills & Shifts  \\
\hline
$\Delta^{\!(\mf{He})} \mathcal{R}_\mf{He}$ & 0\dg  & 
$(3 \Gamma_0 - 2\kappa_\mf{XeHe})/9 \Gamma_0$ & 0.411$\pm$0.008 & $\kappa_\mf{XeHe}$ = -0.0080$\pm$0.0038 & 4 & 34 \\ 
$\Delta^\mf{(Xe)} \mathcal{R}_\mf{He}$ & 0\dg  & 
$(2\kappa_\mf{HeXe} - 3\Gamma_0)/( 2\kappa_\mf{HeXe} + 6 \Gamma_0) $&
-0.639$\pm$0.030 & $\kappa_\mf{HeXe} = $ -0.0059$\pm$0.0009 & 3 & 31  \\
$\Delta^\mf{(Xe)} \mathcal{R}_\mf{Xe}$& 0\dg  & 
$(3 \Gamma_0 - 2\kappa_\mf{HeXe})/9 \Gamma_0$ &
0.369$\pm$0.020 & $\kappa_\mf{HeXe}$ =  -0.0036$\pm$0.0039 & 3 & 31  \\
 $\Delta^{\!(\mf{He})} \mathcal{R}_\mf{Xe}$ & 0\dg  & 
$(2\kappa_\mf{XeHe} - 3\Gamma_0)/(2\kappa_\mf{XeHe} + 6 \Gamma_0)$ &
-0.687$\pm$0.021 & $\kappa_\mf{XeHe} = $ -0.0077$\pm$0.0008 & 4 & 34  \\
$\Delta^{\!(\mf{He})} \mathcal{R}_\mf{He}$ & 90\dg  &
$(1.5 \Gamma_0 + 2\kappa_\mf{XeHe})/4.5 \Gamma_0$ &
 0.140$\pm$0.005 & $\kappa_\mf{XeHe}$ = -0.0010$\pm$0.0006  & 2 & 15 \\
$\Delta^{\!(\mf{Xe})} \mathcal{R}_\mf{He}$ & 90\dg  &
$(2\kappa_\mf{HeXe} + 1.5\Gamma_0)/(2\kappa_\mf{HeXe}-3 \Gamma_0)$ &
 -0.143$\pm$0.056 &$\kappa_\mf{HeXe}$ =  -0.0108$\pm$0.0018  & 2 & 14 \\
$\Delta^{\!(\mf{Xe})} \mathcal{R}_\mf{Xe}$ & 90\dg  &
$(1.5 \Gamma_0 + 2\kappa_\mf{HeXe})/4.5 \Gamma_0$ 
& 0.132$\pm$0.046 & $\kappa_\mf{HeXe}$ = -0.0104$\pm$0.0022 & 2 & 14 \\
$\Delta^{\!(\mf{He})} \mathcal{R}_\mf{Xe}$ & 90\dg  &
$(2\kappa_\mf{XeHe} + 1.5\Gamma_0)/(2\kappa_\mf{XeHe}-3 \Gamma_0)$ &
 -0.157$\pm$0.008 & $\kappa_\mf{XeHe}$ = -0.0102$\pm$0.0009 & 2 & 15 
\label{tab: results}
\end{tabular}
\end{center}
\label{default}
\end{table*}%

\end{document}